\documentclass[aps,preprint,showpacs,preprintnumbers,eqsecnum,amsmath,amssymb]{revtex4}
\usepackage{graphics,epsfig}
\usepackage{graphicx}
\usepackage{dcolumn}
\usepackage{bm}
\begin{document}

\title{Back-reaction of black hole radiation from  Hamilton-Jacobi method}

\author{Chikun Ding }
\email{dingchikun@163.com} \affiliation{Department of Physics and Information
Engineering, Hunan University of Humanities, Science and Technology, Loudi,
Hunan 417000, P. R. China}

 \baselineskip=0.65 cm

\vspace*{0.2cm}

\begin{abstract}
In the frame of Hamilton-Jacobi method, the back-reactions of the radiating
particles together with the total entropy change of the whole system are
investigated. The emission probability from this process is found to be
equivalent to the null geodesic method. However its physical picture is more
clear: the negative energy one of a virtual particle pair is absorbed by the
black hole, resulting in the temperature, electric potential and angular
velocity increase; then the black hole amount of heat, electric charge and
angular momentum can spontaneously transfer to the positive energy particle;
when obtaining enough energy, it can escape away to infinity, visible to
distant observers. And this method can be applied to any sort of horizons and
particles without a specific choice of (regular-across-the-horizon)  coordinates.

\end{abstract}
 \vspace*{1.5cm}

\maketitle

\section{introduction}
Using the quantum field theory in curved spacetime, Hawking \cite{hawking1}
found that the collapsing black hole will lead, at late times, to a radiation
of particles in all modes of the quantum field, with characteristic thermal
spectrum at a temperature $1/8\pi M$. It is generally believed that the pair
productions occur inside and outside the horizon of the black hole and tunnel
across the horizon. In the late time, with knowledge of Feyman's
\cite{feyman} path integral method in the quantum mechanics, he found
\cite{hawking2} that the probability of emission particles from the past
horizon is not the same as the probability of absorption into the future
horizon. The ratio for the Schwarzschild black hole between them is of the
form $\label{ts} \Gamma_{out}=e^{- E/8\pi M}\Gamma_{in}$; for
Reissner-Nordstr\"{o}m black hole is $\label{trn}
\Gamma_{out}=e^{-(E-qV_+)/T_R}\Gamma_{in}$; for Kerr black hole is
$\label{tk} \Gamma_{out}=e^{-(E-m\Omega_+)/T_K}\Gamma_{in}$. These discovers
have excited a lot of interest
\cite{sri,sh1,sh2,ag,man,man1,wil1,wil2,wil3,wil4,berger,medved,chendeyou,
jingyi,jiang,zz1,zz, par,par1,par2,arz,sqw,sqw1,sqw2,ajm,pm1,pm}.

Being enlightened by path-integral method, K. Srinivasan {\it et al}
\cite{sri,sh1,sh2} used Landau's \cite{landau} complex paths method to deduce
radiance without using the Kruskal extension. They treated the radiance as
tunnelling across the singularity and the wave functions as semiclassical
approximation modes $\exp[\frac{i}{\hbar}I(r,t)]$, where $I$ is the classical
action function which can be expanded by $\hbar/i$. To the lowest order, $I$
satisfies the relativistic Hamilton-Jacobi equation which gives a solution
$I_\pm=-Et\pm W(r)+J(x^i)$, where ``$+$" is of outgoing particles and ``$-$"
of incoming particles. $I$ has a pole at horizon $r=r_+$ and the
probabilities of the particles $\Gamma_{out}\sim e^{-2\text{Im}I_+},~~
\Gamma_{in}\sim e^{-2\text{Im}I_-}$.

This complex-path method has been known as Hamilton-Jacobi method after
developed by Angheben {\it et al} \cite{ag} and man {\it et al}
\cite{man,man1}. To ensure the probability is normalized, they used the
boundary conditions for incoming particles which fall behind the horizon
along classically permitted trajectories, i.e. $I=-Et+W(r)+J(x^i)+K$, where
$K$ is a complex normalizing constant. So the total probability is
\begin{eqnarray}\label{gamma2} \Gamma=\Gamma_{out}\sim
e^{-2[\text{Im}I_+-\text{Im}I_-]}.
\end{eqnarray}

However this method can only be used under the condition that the background
spacetime is considered fixed in which the energy conservation is not
enforced during the emission process. Some efforts on extension this method
to dynamic geometry have been done \cite{medved,chendeyou}. But these
generalizations have some crudeness.

Using usual thermodynamic way, we divide the emission process into many
infinite small segments, every one of which can be treated as a quasi-static
process and, the background spacetime can be treated as fixed, there exists
equilibrium temperature. Thus in every segment we can use Hamilton-Jacobi
method to handle. In different segment the instantaneous event horizon is
different. We obtain each action $I_i$ in every tiny time piece after the
particle tunneled across the instantaneous horizon. To get the last action
$I$, the change $\Delta I_i$ between the instantaneous actions should be
considered. Then $I=\sum \Delta I_i\sim \int dI$. After integrating over the
whole process, we obtain the thermal spectrum incorporating the effects of
the back-reaction on the background spacetime, which is the same as that
obtained by the null geodesic method proposed in \cite{wil1}.

There are two different approaches that are used to model tunneling process.
The first method developed was the null geodesic method used by Parikh and
Wilczek\cite{wil1}. Another one is the Hamilton-Jacobi method. There have a
couple of unpleasant features in the null geodesic method: (i) it strongly
relies on a very specific choice of (regular-across-the-horizon) coordinates,
and (ii) it turns upside down the relationship between Hawking's radiation
and back-reaction \cite{vanzo}. The Hamilton-Jacobi method can cope with both
issues. Therefore our method can be applied to any sort of horizons and
particles.

We also study the change of total entropy of the system including black hole
and radiating particles by investigating that where the particle energy comes
from. Then we know why the entropy change of the black hole can be obtain by
probing its radiating particles. Our result is that the difference of total
entropy $\Delta S$ is very small but nonzero, which has some difference from
ref. \cite{wil1} in which the difference of total entropy $\Delta S=0$. As in
ref. \cite{jingyi}, the authors argue that the null geodesic method is only
suitable for the reversible process and the factual emission process is
irreversible which is possible to lose information. Here we argue that
Hamilton-Jacobi method can be suitable for the irreversible process and there
are very few information lost in the emitting process.

 Our paper is outlined as follows. Section
II is for neutral scalar particles radiation from Reissner-Nordstr\"{o}m
black hole and the entropy change of the black hole and radiating particles.
Section III is devoted for charged scalar particles radiating from
Reissner-Nordstr\"{o}m black hole. Section IV is used discussing the charged
scalar particles radiating from Kerr-Newman black hole.
 In section V, charged Dirac particles radiation from Kerr-Newman
 black hole is investigated. Section VI is a summary.

\section{Hawking radiation for Reissner-Nordstr\"{o}m black hole from neutral scalar particles radiation}

The line element for the charged Reissner-Nordstrom back hole is described by
\begin{eqnarray}\label{rn}
ds^2=-\frac{\Delta(r)}{r^2}dt^2+\frac{r^2}{\Delta(r)}dr^2+r^2(d\theta^2+\sin^2\theta
d\varphi^2),
   \end{eqnarray}where $\Delta(r)=r^2-2Mr+Q^2$. When a neutral particle
with the energy $E$ tunnels out across the horizon, the black hole mass and
electric charge $M$ would be decreased to $M-E$ due to energy conservation,
that is to say, the background spacetime is affected by the back-reaction of
the emitting particle, $g^{\mu\nu}(r(M))\rightarrow g^{\mu\nu}(r(M-E))$.
However, because of quantum uncertainty principle, it seems too crude that
the approximation of a discontinuous jump from $M$ to $M-E$ for the black
hole mass. Rather, it will require a ``gradual" transition (relative to
whatever time scale is characteristic of the radiation process). Therefore we
divide this process into many infinite small segments, during $i$th one of
which the particle obtains energy $\Delta\omega_i$, where
$\Delta\omega_i=\omega_i-\omega_{i-1}\ll E$. These segments can be treat as
many quasi-static processes and can be handled by Hamilton-Jacobi method.

A particle of instantaneous energy $\omega_i$ will effectively ``sees" a
spacetime metric of the form
\begin{eqnarray}\label{backreaction}
ds^2=-\frac{\Delta(r(M-\omega_i))}{r^2}dt^2+\frac{r^2}
{\Delta(r(M-\omega_i))}dr^2+r^2(d\theta^2+\sin^2\theta d\varphi^2).
   \end{eqnarray}
In the following subsection, we use Hamilton-Jacobi method to study Hawking
radiation incorporating back-reaction as tunneling, then the entropy change
of the whole system of black hole and radiating particles in the next
subsection.

\subsection{The tunneling process}
 Now we divide the tunneling time $t$ into infinite small pieces $t_i$ and
 use Hamilton-Jacobi method to study these infinite small
processes. The WKB approximation wave function is
\begin{eqnarray}\label{wave}
\phi(t_i,r,\theta,\varphi)=\exp\Big[\frac{i}{\hbar}I_i(t_i,r,\theta,\varphi)
+I'_1(t_i,r,\theta,\varphi) +\mathcal{O}(\hbar)\Big].
   \end{eqnarray}The Klein-Gordon equation is
\begin{eqnarray}\label{kg}
&&\frac{1}
{\sqrt{-g(r(M-\omega_i))}}\partial_{\mu}\Big(\sqrt{-g(r(M-\omega_i))}
\;g^{\mu\nu}(r(M-\omega_i))
\partial_{\nu}\phi\Big)-\frac{u^2}{\hbar^2}\phi =0.
   \end{eqnarray}
   Substituting Eq. (\ref{wave}) into (\ref{kg}), to leading order in $\hbar$, one
get the following relativistic Hamilton-Jacobi equation
   \begin{eqnarray}\label{hj}
g^{\mu\nu}(r(M-\omega_i))(\partial_{\mu} I_i\partial_{\nu} I_i)+u^2=0,
   \end{eqnarray}where there exists a solution in the form
    \begin{eqnarray}I_i=-\omega_i t_i+W_i(r)+J_i(\theta,
   \varphi)+K_i\;,\end{eqnarray}
   where $K_i$ is a complex
   constant normalizing the action function.
 Substituting Eq. (\ref{backreaction}) into
   (\ref{hj}) yields
   \begin{eqnarray}
 W_{i\pm}(r)&=&\pm \int\frac{r^2dr}{\Delta(r(M-\omega_i))}
\sqrt{\omega_i^2-\frac{\Delta(r(M-\omega_i))}{r^2}(u^2+g^{pk}J_pJ_k)}\;,
\nonumber
\end{eqnarray}where $J_p=\partial_p I_i$, $p=\theta,\varphi$.
   One solution of above
    corresponds to the scalar particles moving
away from the black hole (i.e. ``+" outgoing) and the other solution
corresponds to particles moving toward the black hole (i.e. ``$-$" incoming).
Imaginary parts of the action can only come from the vicinity of the pole at
the horizon. Integrating around the pole, the imaginary parts are
\begin{eqnarray}
\text{Im}I_{i\pm}=\text{Im}W_{i\pm}(r)= \pm
f(M-\omega_i)\;\omega_i+\text{Im}K_i\;,\;\;f(M-\omega_i)=\frac{\pi
r_+'^2(M-\omega_i)} {2\sqrt{(M-\omega_i)^2-Q^2}}, \nonumber\\
   \end{eqnarray}
 where $r'_+(M-\omega_i)=M-\omega_i
+\sqrt{(M-\omega_i)^2-Q^2}$ is the instantaneous horizon. Therefore the
action of the particle tunneled across the $i$th instantaneous horizon is
\begin{eqnarray}I_i=-\omega_i t_i+if(M-\omega_i)\;\omega_i+K_i\;.\end{eqnarray}

 When
the energy of the particle gradually approaches to $E$, its action is $I$. To
obtain it, we should consider the change between the $i$th and $(i-1)$th
instantaneous actions $\Delta I_i$
\begin{eqnarray}\Delta I_i=-\Delta\omega_i\; t_i+if(M-\omega_i)\;\Delta\omega_i
+\Delta K_i\;.\end{eqnarray} Then
\begin{eqnarray}I=\sum \Delta I_i=\int_0^E-td\omega+if(M-\omega)\;d\omega
+K\;.\end{eqnarray}
 At last the imaginary part of its action can be
obtained by
\begin{eqnarray}\label{integral}\text{Im}I=\int^{E}_{0}
\frac{\pi r_+'^2} {2\sqrt{(M-\omega)^2-Q^2}}d\omega+\text{Im}K.\end{eqnarray}
 The imaginary
parts of the action of tunneled particle are
\begin{eqnarray}
\text{Im}I_\pm
&=&\pm\int^{E}_{0}\frac{\pi(M-\omega+\sqrt{(M-\omega)^2-Q^2})^2}
{2\sqrt{(M-\omega)^2-Q^2}}d\omega+\text{Im} K \nonumber\\
&=&\pm \frac{\pi}{2}\big[2E(M-\frac{E}{2})-(M-E)\sqrt{(M-E)^2-Q^2}
+M\sqrt{M^2-Q^2}\;\big]+\text{Im} K.\nonumber\\
   \end{eqnarray}

 Using Eq.
(\ref{gamma2}), the emission probability is
\begin{eqnarray}\label{bhentropy}
 \Gamma
=e^{-2\pi\big(2E(M-\frac{E}{2})-(M-E)\sqrt{(M-E)^2-Q^2}
+M\sqrt{M^2-Q^2}\;\big)}=e^{\Delta S_{B-H}},
   \end{eqnarray}
where $\Delta S_{B-H}$ is the difference of Bekenstein-Hawking entropy of the
black hole. This result is the same as in \cite{wil1}. In the following
subsection, we study the entropy change of the whole system and why the
emission probability (\ref{bhentropy}) is related to the entropy change of
the black hole $\Delta S_{B-H}$.

\subsection{The total entropy change of the whole system}
How does the particle obtain energy? If we consider the black hole and its
radiation as a isolated system, then the particle energy can only come from
absorbing black hole's amount of heat $\widetilde{Q}$. Because of vacuum
fluctuation near the event horizon, at sometime the black hole temperature
approaches to $T'(M-\omega)$, which is higher than the particle temperature
$T'(M)$, where $\omega$ is the energy from vacuum fluctuation. So that amount of heat
 can spontaneously transfers from black hole to the particle.

  In the first
  segment, particle energy $0\rightarrow\omega_1$, the
black hole entropy decreases,
 \begin{eqnarray}\Delta
S'_1=-\widetilde{Q}_1/T'(M-\omega_1),\end{eqnarray}
 where $S'$ is the black hole
 entropy, $T'(M-\omega_1)=\frac{1}{2\pi}\sqrt{(M-\omega_1)^2-Q^2}
 /r_+'^2(M-\omega_1)$. The particle entropy $S''$ increases,
\begin{eqnarray}\Delta S''_1=\widetilde{Q}_1/T'(M),\end{eqnarray}
and the entropy of the system increase
\begin{eqnarray}\Delta
S_1=\Delta S''_1+\Delta
S'_1=\widetilde{Q}_1/T'(M)-\widetilde{Q}_1/T'(M-\omega_1)>0.\end{eqnarray} It
shows that the radiating process is irreversible and $\omega_1$ is
 \begin{eqnarray}\Delta\omega_1=\omega_1-0=-T'(M-\omega_1)\Delta S'_1\;.\end{eqnarray}
 It is easy to see that we can really obtain the entropy change of the black
 hole by probing its radiating particle energy and, the
emission probability (\ref{bhentropy}) is really related to the entropy
change of the black hole $\Delta S_{B-H}$. At the end of the first segment,
the particle temperature approached to $T'(M-\omega_1)$ and,  for the black
hole, it approached to $T'(M-\omega_2)$.

In the second segment, particle energy $\omega_1\rightarrow\omega_2$, the
particle absorbs heat $\Delta\omega_2=\omega_2-\omega_1=-T'(M-\omega_2)\Delta
S'_2$, and the increase of the system entropy is
\begin{eqnarray}\Delta
S_2=\Delta S''_2+\Delta S'_2=\Delta\omega_2/T'(M-\omega_1)-\Delta\omega_2
/T'(M-\omega_2).\end{eqnarray}

When the energy of the particle has approached to $E$, the black hole
temperature
 is $T(M-E)$,
and the total increase of the system entropy is
\begin{eqnarray}\label{entropy}\Delta
S=\Delta S_1+\Delta
S_2+\cdots=\Delta\omega/T'(M)-\Delta\omega/T'(M-E)<\Delta\omega/T'(M)\end{eqnarray}
under condition that $\Delta\omega_1=\Delta\omega_2=\cdots=\Delta\omega$. Due
to $\Delta\omega\ll E$, the total increase of the system entropy is very
small but nonzero. It is the same as in Ref. \cite{zurek}, but has some
difference from Ref. \cite{wil1} in which $\Delta S=0$ ({\it In the limit the
emitted particle carries away the entire mass and charge of the black hole,
there are exp$(S_{B-H})$ states in total and the probability of findings a
shell containing all of the mass of the black hole is proportional to
exp$(-S_{B-H})$}).

\section{Hawking radiation for Reissner-Nordstr\"{o}m black hole from charged
 scalar particles radiation}

A particle of instantaneous energy $\omega_i$ and charge $q_i'$ will
effectively sees a spacetime metric of the form
\begin{eqnarray}\label{charged}
ds^2=-\frac{\Delta(r(M-\omega_i,Q-q_i'))}{r^2}dt^2+\frac{r^2}
{\Delta(r(M-\omega_i,Q-q_i'))}dr^2+r^2(d\theta^2+\sin^2\theta d\varphi^2).
\nonumber\\
   \end{eqnarray}

The charged Klein-Gordon equation is
\begin{eqnarray}\label{kg2}
&&\frac{(\partial_{\mu}-iq_i'A_\mu)\Big(\sqrt{-g(r(M-\omega_i,Q-q_i'))}
\;g^{\mu\nu}(r(M-\omega_i,Q-q_i')) (\partial_{\nu}-iq_i'A_\nu)\phi\Big)}
{\sqrt{-g(r(M-\omega_i,Q-q_i'))}} -\frac{u^2}{\hbar^2}\phi=0,\nonumber\\
   \end{eqnarray}
where $A_{{\mu}}=(-(Q-q_i')/r,0,0,0)$ is the potential of the electromagnetic
field of
   the background spacetime.
   Substituting Eq. (\ref{wave}) into (\ref{kg2}), to leading order in $\hbar$, one
get the following relativistic Hamilton-Jacobi equation
   \begin{eqnarray}\label{hj2}
g^{\mu\nu}(r(M-\omega_i,Q-q_i'))(\partial_{\mu} I_i\partial_{\nu}
I_i+q_i'^2A_\mu A_\nu-2q_i'A_\mu\partial_\nu I_i)+u^2=0.
   \end{eqnarray}
 Substituting Eq. (\ref{charged}) into
   (\ref{hj2}) yields
   \begin{eqnarray}
 W_{i\pm}(r)&=&\pm \int\frac{r^2dr}{\Delta(r(M-\omega_i,Q-q_i'))}
\sqrt{\big[\omega_i-\frac{q_i'(Q-q_i')}{r}\big]^2
-\frac{\Delta(r(M-\omega_i,Q-q_i'))}{r^2}(u^2+g^{pk}J_pJ_k)}\;. \nonumber
\end{eqnarray}
   Integrating around the pole, the imaginary parts are
\begin{eqnarray}
\text{Im}I_{i\pm}=\text{Im}W_{i\pm}(r)= \pm\frac{\pi
r_+'^2(M-\omega_i,\;Q-q_i')}
{2\sqrt{(M-\omega_i)^2-(Q-q_i')^2}}\Big(\omega_i-q_i'V'_+(M-\omega_i,\;Q-q_i')\Big)
+\text{Im}K,\nonumber\\
   \end{eqnarray}
 where $V'_+(M-\omega_i,\;Q-q_i')=(Q-q_i')/r'_+(M-\omega_i,\;Q-q_i')$ is the instantaneous
 electric potential near the outer horizon,
  and $r'_+(M-\omega_i,\;Q-q_i')=M-\omega_i
+\sqrt{(M-\omega_i)^2-(Q-q_i')^2}$ is instantaneous the horizon. The action
of the particle tunneled across the $i$th instantaneous horizon is
\begin{eqnarray}I_i=-\omega_i t_i+i\frac{\pi r_+'^2}
{2\sqrt{(M-\omega_i)^2-(Q-q_i')^2}}(\omega_i-q_i'V'_+)+K_i\;,\end{eqnarray}
and the change between the $i$th and $(i-1)$th instantaneous actions $\Delta
I_i$
\begin{eqnarray}\Delta I_i=-\Delta\omega_i\; t_i+i\frac{\pi r_+'^2}
{2\sqrt{(M-\omega_i)^2-(Q-q_i')^2}}(\Delta\omega_i-\Delta q_i'V'_+) +\Delta
K_i\;,\end{eqnarray}where $\Delta q_i'=q_i'-q'_{i-1}$.
 When
the energy of the particle gradually approaches to $E$, its action is $I$
\begin{eqnarray}\label{integral4}I=\sum \Delta I_i=\int_0^E-td\omega
+i\int_{(0,0)}^{(E,q)}\frac{\pi r_+'^2}
{2\sqrt{(M-\omega_i)^2-(Q-q_i')^2}}(d\omega-dq'V'_+) +K\;.\nonumber\\
\end{eqnarray}
 Using Eq. (\ref{integral4}), the imaginary
parts of the action of tunneled particle are
\begin{eqnarray}
\text{Im}I_\pm
&=&\pm\int^{(E,q)}_{(0,0)}\frac{\pi(M-\omega+\sqrt{(M-\omega)^2-(Q-q')^2})^2}
{2\sqrt{(M-\omega)^2-(Q-q')^2}}(d\omega-V_+'dq')+\text{Im} K \nonumber\\
&=&\pm \frac{\pi}{2}\big[2E(M-\frac{E}{2})-(M-E)\sqrt{(M-E)^2-(Q-q)^2}
\nonumber\\
&&+M\sqrt{M^2-(Q-q)^2}-Qq+\frac{1}{2}q^2\;\big]+\text{Im} K.
   \end{eqnarray}

 Using Eq.
(\ref{gamma2}), the emission probability is
\begin{eqnarray}
 \Gamma
=e^{-2\pi\big(2E(M-\frac{E}{2})-(M-E)\sqrt{(M-E)^2-(Q-q)^2}
+M\sqrt{M^2-(Q-q)^2}-Qq+\frac{1}{2}q^2\;\big)}=e^{\Delta S_{B-H}},\nonumber\\
   \end{eqnarray}
which is the same as in \cite{zz1}.

Now turn to the acquisition of particle energy. At sometime, the black hole
temperature rises from $T'(M,\;Q)$ to $T'(M-\omega_1,\; Q-q_1')$, its horizon
shrinks from $r_+'(M,\;Q)$ to $r_+'(M-\omega_1,\; Q-q_1')$, which lead to
increase of the electric potential of the black hole, $V_+'(M,\;Q)\rightarrow
V_+'(M-\omega_1,\; Q-q_1')$, where $T'(M-\omega_1,\;
Q-q_1')=\frac{1}{2\pi}\sqrt{(M-\omega_1)^2-(Q-q_1')^2}/
r_+'^2(M-\omega_1,\;Q-q_1')$. Then the amount of heat and electric charge
flow to the particle, which will lead to another further increase of black
hole temperature and electric potential. The particle energy comes from two
ways, absorbing black hole's internal energy and electric potential energy
$q'_iV'_+$. Energy $\omega_i$ is
 \begin{eqnarray}\Delta\omega_i=\omega_i-\omega_{i-1}=-
 T'(M-\omega_i)\Delta S_i'+V'_+\Delta q_i'\;,\end{eqnarray}
 where $\Delta q_i'=q_i'-q'_{i-1}$.

\section{Hawking radiation for Kerr-Newman black hole from charged
scalar particles radiation } The ``no hair" theorem stated that all
information about the collapsing body was lost from the outside region apart
from three conserved quantities: the mass, the angular momentum and the
electric charge, the final state of most of collapsing star is Kerr-Newman
black hole. In the Boyer-Lindquist coordinate, its line element in four
dimensional spacetime is described by
\begin{eqnarray}\label{kn}
&&ds^2=-\left(1-\frac{2Mr}{\rho^2}\right)dt^2-\frac{4Mra\sin^2\theta}
{\rho^2}dtd\varphi+\frac{\rho^2}{\triangle}dr^2\nonumber\\
&&+\rho^2d\theta^2+\left(r^2+a^2+\frac{2Mra^2\sin^2\theta}{\rho^2}\right)\sin^2\theta
d\varphi^2,
   \end{eqnarray}with
   \begin{eqnarray}\nonumber
   &&\rho^2=r^2+a^2\cos^2\theta,~~~~ \triangle=r^2-2Mr+a^2+Q^2=(r-r_+)(r-r_-)\nonumber\\
   &&r_+=M+\sqrt{M^2-a^2-Q^2},~~~~r_-=M-\sqrt{M^2-a^2-Q^2},
\end{eqnarray}where $M$ is the mass of the black hole and $a=J/M$ is the angular momentum
parameter; $r_-$ and  $r_+$ are the inner and event horizons; its
electromagnetic field potential is
$A_{{\mu}}=(-Qr/\rho^2,~0,~0,~Qra\sin^2\theta/\rho^2)$. When a particle with
the energy $E$, electric charge $q$ and angular momentum $j$ tunnels out
across the horizon, the black hole mass, angular momentum, and electric
charge $M,\;Q,\;J$ would be decreased to $M-E,\;Q-q,\;J-j$ due to energy,
electric charge and angular momentum conservation.
 We also divide this process into many infinite small
segments, during the $i$th one of which the particle obtains energy
$\Delta\omega_i$, angular momentum $\Delta j'_i$, where $\Delta
j'_i=j'_i-j'_{i-1}$. These segments can be treat as many quasi-static
processes and can be handled by Hamilton-Jacobi method.

A particle of instantaneous energy $\omega_i$, charge $q_i'$ and angular
momentum $j'_i$ will effectively sees a spacetime metric of the form
\begin{eqnarray}\label{knp}
&&ds^2=-\left(1-\frac{2(M-\omega_i)r}{\tilde{\rho}^2}\right)dt^2
-\frac{4(M-\omega_i)r\tilde{a}_i\sin^2\theta}
{\tilde{\rho}^2}dtd\varphi+\frac{\tilde{\rho}^2}{\triangle(r(M-\omega_i,
Q-q'_i,J-j_i'))}dr^2\nonumber\\
&&+\tilde{\rho}^2d\theta^2+\left(r^2+\tilde{a}_i^2+\frac{2(M-\omega_i)r\tilde{a}_i^2
\sin^2\theta}{\tilde{\rho}^2}\right)\sin^2\theta d\varphi^2,
   \end{eqnarray}where $\tilde{a}_i=(J-j_i')/(M-\omega_i)$,
   $\tilde{\rho}^2=r^2+\tilde{a}_i^2\cos^2\theta$.

We divide tunneling time $t$, rotating angle $\varphi$ into infinite small
pieces $t_i,\;\varphi_i$ and use Hamilton-Jacobi method. Due to the energy
and angular momentum conservation, its instantaneously action function must
be of the form
\begin{eqnarray}I_i=-\omega_i t_i+W_i(r)+j_i'\varphi_i+\tilde{J_i}(\theta)+K_i.\end{eqnarray} Taking the line
element (\ref{knp}) into Hamilton-Jacobi equation (\ref{hj2}), one find
\begin{eqnarray}\label{separation}
&&\triangle^2W_i'^2(r)+\Big[\triangle u^2r^2-\tilde{a}_i^2 \widetilde{j_i'}^2
+4(M-\omega_i )r\tilde{a}_i \widetilde{j_i'}\widetilde{\omega}_i
-(r^2+\tilde{a}_i^2)^2\widetilde{\omega}_i^2\Big]+\triangle\lambda=0,\nonumber\\
&&~~~~\lambda=\Big[\tilde{a}_i^2\sin^2\theta
\widetilde{\omega}_i^2+\tilde{J_i}'^2(\theta)+\frac{ \widetilde{j_i'}^2}
{\sin^2\theta}+u^2\tilde{a}_i^2\cos^2\theta\Big],
   \end{eqnarray}
 where \begin{eqnarray}
   \widetilde{\omega}_i=\omega_i-\frac{q_i'(Q-q_i')r}{\tilde{\rho}^2},
   ~~\widetilde{j_i'}=j'-\frac{q_i'(Q-q_i')r}{\tilde{\rho}^2}\tilde{a}_i\sin^2\theta.
   \end{eqnarray}
   Then the imaginary parts of the action function are
\begin{eqnarray}\label{ww3}
 \text{Im}I_{i\pm}&=&\pm  \text{Im}\left[\int dr
\frac{1}{\triangle}\sqrt{(r^2+\tilde{a}_i^2)^2\big(\widetilde{\omega}_i
-\frac{\widetilde{j_i'}\tilde{a}_i}{r^2+\tilde{a}_i^2}\big)^2
-\triangle(u^2r^2+\lambda-2\widetilde{j_i'}\tilde{a}_i\widetilde{\omega}_i)}
\;\right]+\text{Im}K_i\nonumber\\
&=&\pi\frac{r'^{2}_++\tilde{a}_i^2}{2(r_+'-M+\omega_i)}
\left(\widetilde{\omega}_{i+}
-\frac{\widetilde{j_i'}_+\tilde{a}_i}{r^{'2}_++\tilde{a}_i^2}\right)+\text{Im}K_i
\nonumber\\
&=&\pi\frac{r'^{2}_++\tilde{a}_i^2}{2(r_+'-M+\omega_i)}\left(\omega_i
-\frac{q_i'(Q-q_i')r'_+}{r'^{2}_++\tilde{a}_i^2}
-\frac{j_i'\tilde{a}_i}{r'^{2}_++\tilde{a}_i^2}\right)+\text{Im}K_i\nonumber\\
&=&\pi\frac{r'^{2}_++\tilde{a}_i^2}{2(r_+'-M+\omega_i)}\left(\omega_i-q_i'V'_+
-j_i'\Omega'_+\right)+\text{Im}K_i,
   \end{eqnarray}
where $r'_+(M-\omega_i,\;Q-q_i',\;J-j'_i)=M-\omega_i+\sqrt{(M-\omega_i)^2
   -(Q-q_i')^2-\tilde{a}_i^2}$, $V'_+(M-\omega_i,\;Q-q_i',\;J-j'_i)
   =(Q-q_i')r'_+/(r'^{2}_++\tilde{a}_i^2)$
   is the electromagnetic potential on the horizon,
   $\Omega'_+(M-\omega_i,\;Q-q_i',\;J-j'_i)=\tilde{a}_i/(r'^{2}_++\tilde{a}_i^2)$
   is the dragging velocity of the horizon.

When the energy, charge and angular momentum of the particle gradually
approaches to $E,\;q,\;j$, the imaginary part of its action can be obtained
by
\begin{eqnarray}\label{integral2}\text{Im}I=\int^{(E,q,j)}_{(0,0,0)}
\pi\frac{r'^{2}_++\tilde{a}_i^2}{2(r_+'-M+\omega)}(d\omega-V'_+dq'
-\Omega_+'dj')+\text{Im}K.\end{eqnarray}
 Using Eq. (\ref{integral2}), the imaginary
parts of the action of tunnelled particle are
\begin{eqnarray}
\text{Im}I_\pm &=&\pm
\frac{\pi}{2}\big[2E(M-\frac{E}{2})-(M-E)\sqrt{(M-E)^2-(Q-q)^2-\tilde{a}^2}
\nonumber\\
&&+M\sqrt{M^2-(Q-q)^2-a^2}-Qq+\frac{1}{2}q^2+\tilde{a}^2-a^2\;\big]
+\text{Im} K,
   \end{eqnarray}where $\tilde{a}=(J-j)/(M-E)$.

Using Eq. (\ref{gamma2}), the emission probability is
\begin{eqnarray}
 \Gamma
=e^{-2\pi\big(2E(M-\frac{E}{2})-(M-E)\sqrt{(M-E)^2-(Q-q)^2-\tilde{a}^2}
+M\sqrt{M^2-(Q-q)^2-a^2}-Qq+\frac{1}{2}q^2+\tilde{a}^2-a^2\;\big)}=e^{\Delta
S_{B-H}}.\nonumber\
   \end{eqnarray}
If set $j=Ea$ for special case, then $\tilde{a}=a$, so this result is the
same as in \cite{zz,jiang}.

Now turn to the acquisition of particle energy near the horizon of
Kerr-Newman black hole. At sometime, the black hole temperature rises from
$T'(M,\;Q,\;J)$ to $T'(M-\omega_1,\; Q-q_1',\;J-j'_1)$, its horizon shrinks
from $r_+'(M,\;Q,\;J)$ to $r_+'(M-\omega_1,\; Q-q_1',\;J-j'_1)$, which lead
to increase of the electric potential and the angular velocity of the black
hole, $V_+'(M,\;Q,\;J)\rightarrow V_+'(M-\omega_1,\; Q-q_1',\;J-j'_1)$,
$\Omega_+'(M,\;Q,\;J)\rightarrow \Omega_+'(M-\omega_1,\; Q-q_1',\;J-j'_1)$,
where $T'(M-\omega_1,\;
Q-q_1',\;J-j'_1)=\frac{1}{2\pi}\sqrt{(M-\omega_1)^2-(Q-q_1')^2-\tilde{a}_1^2}/
(r_+'^2+\tilde{a}_1^2)$. Then the amount of heat, electric charge and angular
momentum flow to the particle, which will lead to another further increase of
black hole temperature, electric potential and angular velocity. The particle
energy comes from three ways, absorbing black hole's internal energy,
electric potential energy $q'_iV'_+$  and rotational kinetic energy
$\Omega_+'\Delta j_i'$
\begin{eqnarray}\Delta\omega_i=-T'(M-\omega_i)\Delta S_i'+V'_+\Delta q_i'
+\Omega_+'\Delta j_i'\;.\end{eqnarray}

\section{Hawking radiation for Kerr-Newman black hole from charged Dirac particles tunnelling }
In this section, we extend the Hamilton-Jacobi method to Dirac field. The
charged Dirac equation is\begin{eqnarray}\label{dirac} &&\left[\gamma^\alpha
e^{{\mu}}_\alpha(\partial_{{\mu}}+\Gamma_{{\mu}}-iq_i'A_{\mu})
+\frac{u}{\hbar}\right] \psi=0,
 \end{eqnarray}
  with
  \begin{eqnarray}
&&\Gamma_{{\mu}}=\frac{1}{8}[\gamma^a,\gamma^b]e^{{\nu}}_ae_{b{{\nu}};{{\mu}}},
  \nonumber  \end{eqnarray}
where $\gamma^a$ are the Dirac matrices and  $e^{{\mu}}_a$ is the
   inverse tetrad defined by
     $\{e_a^{{\mu}}\gamma^a,~~~e_b^{{\nu}}\gamma^b\}=2g^{{{\mu}}{{\nu}}}
 \times1$. For the Kerr-Newman metrics in Boyer-Lindquist coordinate (\ref{knp}), the nonzero tetrad elements can be
     taken as
   \begin{eqnarray}\label{tetrad}
   &&e_0^{t_i}=\frac{1}{\sqrt{1-\frac{2(M-\omega_i)r}{\tilde{\rho}^2}}},
   ~~e^{t_i}_3=-\frac{2(M-\omega_i)r\tilde{a}_i\sin\theta}{\tilde{\rho}
   \sqrt{\triangle^2-\tilde{a}_i^2\sin^2\theta\triangle}},\nonumber\\
   &&e^r_1=\frac{\sqrt{\triangle}}{\tilde{\rho}},~~~~
   e^\theta_2=\frac{1}{\tilde{\rho}},
   ~~e^{\varphi_i}_3=\frac{\sqrt{\triangle-\tilde{a}_i^2\sin^2\theta}}
   {\tilde{\rho}\sin\theta\sqrt{\triangle}},\nonumber\
\end{eqnarray}where $\triangle=r^2+\tilde{a}_i^2-2(M-\omega_i)r+(Q-q_i')^2$.
We employ the following ansatz for the Dirac field
   \begin{eqnarray}\label{psi}
  &&\psi_{i\uparrow}=\bigg(\begin{array}{ccc}A(t_i,r,\theta,\varphi_i)
  \xi_\uparrow\nonumber\\
   B(t_i,r,\theta,\varphi_i)\xi_\uparrow\end{array}\bigg)
   \exp\big(\frac{i}{\hbar}I_{i\uparrow}(t_i,r,\theta,\varphi_i)\big)
   =\left(\begin{array}{ccc}A(t_i,r,\theta,\varphi_i)\nonumber\\ 0\nonumber\\
   B(t_i,r,\theta,\varphi_i)\nonumber\\0\end{array}\right)
   \exp\big(\frac{i}{\hbar}I_{i\uparrow}(t_i,r,\theta,\varphi_i)\big),\nonumber\\
   &&\psi_{i\downarrow}=\bigg(\begin{array}{ccc}C(t_i,r,\theta,\varphi_i)
   \xi_\downarrow\nonumber\\
   D(t_i,r,\theta,\varphi_i)\xi_\downarrow\end{array}\bigg)
   \exp\big(\frac{i}{\hbar}I_{i\downarrow}(t_i,r,\theta,\varphi_i)\big)
   =\left(\begin{array}{ccc}0\nonumber\\ C(t_i,r,\theta,\varphi_i)\nonumber\\
   0\\D(t_i,r,\theta,\varphi_i)\nonumber\end{array}\right)
   \exp\big(\frac{i}{\hbar}I_{i\downarrow}(t_i,r,\theta,\varphi_i)\big),\nonumber\\
   \end{eqnarray}
where ``$\uparrow$" and ``$\downarrow$" represent the spin up and spin down
cases, and $\xi_{\uparrow}$ and $\xi_{\downarrow}$ are the eigenvectors of
$\sigma^3$. Inserting Eqs. (\ref{tetrad}), (\ref{psi}) into the Dirac
equation (\ref{dirac}) and employing $I_{i\uparrow}=-\omega_i
t_i+W_i(r)+j_i'\varphi_i+\tilde{J_i}(\theta)+$Im$K_i$, to the lowest order in
$\hbar$ we obtain\begin{eqnarray}\label{aa}
  && -e^{t_i}_0A\Big(\omega_i-\frac{q_i'(Q-q_i')r}{\tilde{\rho}^2}\Big)
  +e^r_1BW_i'(r)
   +u A=0,\\\label{bb}&&e^{t_i}_0B
   \Big(\omega_i-\frac{q_i'(Q-q_i')r}{\tilde{\rho}^2}\Big)-e^r_1AW_i'(r)
   +u B=0, \\
   \label{cc}
  &&
  B\left[-ie^{t_i}_3\Big(\omega_i-\frac{q_i'(Q-q_i')r}{\tilde{\rho}^2}
  \Big)+e^\theta_2\tilde{J_i}'(\theta)+ie^{\varphi_i}_3
  \Big(j_i'-\frac{q_i'(Q-q_i')r}{\tilde{\rho}^2}\tilde{a}_i\sin^2\theta\Big)
  \right]=0,\nonumber\\
  \\\label{dd}&&
  -A\left[-ie^{t_i}_3\Big(\omega_i-\frac{q_i'(Q-q_i')r}{\tilde{\rho}^2}
  \Big)+e^\theta_2\tilde{J_i}'(\theta)+ie^{\varphi_i}_3
  \Big(j_i'-\frac{q_i'(Q-q_i')r}{\tilde{\rho}^2}\tilde{a}_i
  \sin^2\theta\Big)\right]=0,\nonumber\\
\end{eqnarray}
where we consider only the positive frequency contributions
without loss of generality. From above four equations, we can
obtain\begin{eqnarray} &&\frac{\triangle}{\tilde{\rho}^2}W_i'^2(r)+u^2
-\frac{\Big(\omega_i-\frac{q_i'(Q-q_i')r}{\tilde{\rho}^2}\Big)^2}{\rho^2\triangle}
\left[\frac{4(M-\omega_i)^2r^2\tilde{a}_i^2\sin^2\theta}
{\triangle-\tilde{a}_i^2\sin^2\theta}+(r^2+\tilde{a}_i^2)^2-\triangle
\tilde{a}_i^2\sin^2\theta\right]=0,\nonumber\\
&&\frac{1}{\tilde{\rho}^2}\left[\tilde{J_i}'^2(\theta)
+\frac{\Big(j_i'-\frac{q_i'(Q-q_i')r}{\tilde{\rho}^2}\tilde{a}_i
\sin^2\theta\Big)^2
(\triangle-\tilde{a}_i^2\sin^2\theta)}{\triangle\sin^2\theta}
+\frac{4(M-\omega_i)^2r^2\tilde{a}_i^2\Big(\omega_i-\frac{q_i'(Q-q_i')r}
{\tilde{\rho}^2}\Big)^2
\sin^2\theta}{\triangle(\triangle-\tilde{a}_i^2\sin^2\theta)}\right.
\nonumber\\
&&  \left.
+2\Big(j_i'-\frac{q_i'(Q-q_i')r}{\tilde{\rho}^2}\tilde{a}_i\sin^2\theta\Big)
\Big(\omega_i-\frac{q_i'(Q-q_i')r}{\tilde{\rho}^2}\Big)
\frac{2(M-\omega_i)r\tilde{a}_i}{\triangle}\right]=0.
 \end{eqnarray}
 After calculating the above two equations, we find
 \begin{eqnarray}
&&\triangle^2W_i'^2(r)-(r^2+\tilde{a}_i^2)^2\widetilde{\omega}_i^2
-\tilde{a}_i^2 \widetilde{j_i'}^2 +4(M-\omega_i )r\tilde{a}_i
\widetilde{j_i'}\widetilde{\omega}_i\nonumber\\
&&+ \triangle\left[\tilde{J_i}'^2(\theta)+\frac{ \widetilde{j_i'}^2}
{\sin^2\theta}+u^2r^2+u^2\tilde{a}_i^2\cos^2\theta- \tilde{a}_i^2\sin^2\theta
\widetilde{\omega}_i^2\right]=0,\nonumber
 \end{eqnarray}which is the same as the Eq. (\ref{separation}),
   so the Hawking radiation is again
   recovered. The result can be interpreted that the black holes
   radiate different spin weight of particles at the same
   temperature.
\section{summary}

 In
the Hamilton-Jacobi framework, we have naturally discussed
Reissner-Nordstr\"{o}m and Kerr-Newman black holes' radiance  with
back-reaction from neutral scalar, charged scalar and Dirac particles
radiation.
 Hamilton-Jacobi method can only be used to
modeling the case that the background geometry is considered fixed. To handle
the back-reaction of the radiating particles, we should first divide the
radiation time into a series of infinite small pieces in which the small
process can be a quasi-static one. Then using Hamilton-Jacobi method, we obtain
the instantaneous action of the particle tunneled across the instantaneous
black hole horizon. To get the last action, we should find the changes
between the $i$th and $(i-1)$th instantaneous action and sum them. The last
result is the same as that obtained via the null geodesic method.

The physical meaning is obvious in this processing. Due to vacuum fluctuation
near horizon, a virtual particle pair is created. The negative energy
particle is absorbed by the black hole, resulting in the black hole mass
decrease, while temperature, electric potential and angular velocity
increase. Then the black holes' amount of heat, electric charge and angular
momentum can spontaneously transfer to the positive energy particle. This
process also results in the black hole temperature, electric potential and
angular velocity further increase and further transfer. When the particle
obtains enough energy, it can escape away to infinity, visible to distant
observers.

We have also studied the change of total entropy of the system including
black hole and radiating particles and, answered that why the entropy change
of the black hole can be obtain by probing its radiating particles. Our
result is that the difference of total entropy $\Delta S$ is very small and can be ignored.

\section*{Acknowledgments}

This work was supported by the National Natural Science Foundation of China
under No. 11247013; Hunan Provincial NSFC No. 11JJ3014, the Scientific
Research Fund of Hunan Provincial Education Department No. 11B067 and, the
Foundation for the Author of Hunan Provincial Excellent Doctoral Dissertation
No. YB2012B034; Aid program for Science and Technology Innovative Research
Team in Higher Educational Institutions of Hunan Province.


\end{document}